\documentstyle[prb,aps,twocolumn,amssymb,amsfonts,epsfig]{revtex}

\title{Electronic structure and magnetism for FeSi$_{(1-x)}$Ge$_x$ from supercell calculations.}
\author{ 
T. Jarlborg\\ 
DPMC, University of Geneva, \\
24 Quai Ernest Ansermet, CH-1211 Geneva 4, Switzerland \\}

\begin{document}

\maketitle 
\date{\today}
  
\begin{abstract}
Recent studies of FeSi$_{(1-x)}$Ge$_x$, which
found a transition from an insulating to a magnetic metallic state
near $x$=0.25, have revived the discussion about the role of strong correlation
in these systems. Here are
spin polarized band calculations made for 64-atom
supercells of FeSi$_{(1-x)}$Ge$_x$ for different $x$ and
different volumes for large $x$. The results show that the small band gap
in FeSi is closed for $x \geq 0.3$, because of both substitutional
disorder and increased volume. 
Ferromagnetism appears near this composition and becomes enforced for increasing $x$. 
The $x$-dependence of the electronic specific heat
can be understood from the exchange splitting of the
density-of-states near the gap. Strong volume dependencies for the properties of FeGe suggest 
experiments using pressure instead of $x$ for investigations of the gap.

\end{abstract}


\section{Introduction}
\vspace{0.5cm}

The unusual properties of the cubic compound $\epsilon$-FeSi, investigated
early by Jaccarino et al \cite{jac}, continue to be the subject
of many theoretical and experimental studies \cite{man}-\cite{voc}. Band calculations 
based on the local density approximation (LDA) \cite{lda}, showing that FeSi is 
a very narrow band gap semiconductor with a gap of about 6 mRy \cite{mat}-\cite{voc},
can explain many observed low-temperature properties. 
 The sharpness of the density-of-states (DOS) below the gap is observed
as a dispersionless feature in the spectra of angular resolved
photoemission  \cite{par} taken at low temperature, T. 
Optical measurements reveal that the gap is gradually filled
at higher T, and an unusual feature appears to be that the spectral weight
is not conserved \cite{sch,deg}.
The problem is also to understand 
why the material behaves as a metal with large
magnetic susceptibility $\chi(T)$ at higher T.
Electronic excitations within the band structure (determined at T=0), described
by the Fermi-Dirac function, are not able to explain such T-dependent properties \cite{jac,man}. 
 Many theories, mostly based on some form of strong correlation, have been proposed 
for an explanation of the unusual T-dependence
 \cite{mor,app,ani,don}. 
On the other hand, electronic structure calculations within the LDA can explain the T-dependent properties
when the effect of thermal disorder is included  \cite{jar99}. The filling of the gap is an effect 
of structural disorder for T larger than about 150 K, and
the material behaves as a Stoner enhanced paramagnet above this temperature \cite{jar99}.

Different dopings of FeSi or other compositions like FeGe, MnSi, etc.
 lead to a large variety of properties, and metallic magnetism is often
found \cite{mena}.  The
LDA approach has been used to analyze thermoelectric properties in doped FeSi \cite{jar01}. 
Recent studies  \cite{yeo,ani2} focused on
Ge substitutions on Si sites, in which the evolution of the gap and other properties were followed
continuously between the isoelectronic FeSi and FeGe systems. Pure FeGe is metallic and magnetic 
with a long-range spiral spin alignment, and the saturation moment is about 1 $\mu_B$ per Fe \cite{yeo,leb,lun}.
The alloy FeSi$_{(1-x)}$Ge$_x$
shows a transition from an insulating state to a metallic ferromagnetic one near $x$=0.25, with
enforced magnetism at larger $x$ \cite{yeo}. Some, but not all properties are interpreted
in terms of the Kondo insulator model with a Hubbard $U$ as parameter \cite{yeo}. In order
to pursue the search for the alternative solution based on LDA band structures, we study
here the evolution of the gap and the magnetic moment as function 
of Ge substitutions with 
special attention to the effect of substitutional disorder. Methods like the virtual crystal
approximation, which interpolate potential parameters between the pure constituents, are insufficient
to study the effect of disorder. Therefore we use supercells with different site occupations for the
band calculations of the FeSi$_{(1-x)}$Ge$_x$ alloys.

\vspace{0.5cm}
\section{Method of calculation.}
\vspace{0.5cm}

 The bandstructure for the ordered material
can be described by the use of a supercell, although it is a more complicated calculation
than for the normal cell.  The bands of the normal cell are folded back into
the smaller Brillouin Zone of the supercell, giving more bands per k-point. 
The bands are degenerate along different directions in k-space,
but if there is disorder (structural or substitutional) there will be slighly different band dispersions
along different directions. The average energy of a band along say $k_x$ and $k_y$ may be similar
to the energy for the ordered case, but the non-degeneracy along different directions can be interpreted as a
band broadening of the original band. Thus, the band broadening or smearing of the bands near
the gap, which becomes important for the properties of the FeSi-like materials, is because of a
number of non-degenerate states that deviate from the the original energy as the degree of disorder increases. 
These effects, as well as a possible drift of the average energy
due to disorder, are contained in the supercell approach. Disorder of real alloys does not show the 
long-range periodicity of the supercell, but this a minor problem when large supercells can be used.

The spin-polarized version of LDA can be used to determine the magnetic moment, but the question of
spiral order of the spin moments in FeGe is not addressed in this work. The calculated moments will be
compared with saturation moments at high field (0.3 T) at which the moments are aligned. Since a magnetic
field of this amplitude is very small on the energy scales in a band calculation (1 mRy corresponds to 230 T)
it is unlikely that this field can modifiy exchange splitting or the electronic structure.

The self-consistent
Linear Muffin-Tin Orbital method is applied to 64-atom supercells
(8 basic unit cells of the B20-structure) as is described for the calculation
for structurally disordered FeSi \cite{jar99}. 
The linearization energies are taken near the center of each $\ell$-band. 
The LDA potential contains
no special on-site correction due to correlation.
All sites are nonequivalent in the disordered cells. 
The self-consistent iterations use initially 8 k-points in 1/8 of the Brillouin Zone,
while they are terminated by iterations using 27 k-points. The band gaps in the two sets 
of k-points agree to within 0.5 mRy, but the magnetic moments in the metallic cases can differ by 20-30 
percent for some configurations. This difference comes from difference in the DOS near the gap between
the two sets of k-points. A comparison of the DOS from a case using of 27 and 64 k-points show no large
difference and it is expected that the convergence of the moments from 27 k-points is satisfactory.
The calculations include
a thermal smearing of 1 mRy in the electronic occupation (Fermi-Dirac). This stabilizes the
self-consistency, but there is also a physical reason behind the smearing; The zero-point motion of the
atomic positions, leading to some structural disorder, can be estimated to be of the order 0.5 percent of interatomic
distances for appropriate values of force constants and atomic masses \cite{js,voc}. 
From various arguments it can be concluded that an essential effect of band smearing on the electronic structure
near the Fermi energy, $E_F$, can to some extent
be modeled through the Fermi-Dirac occupation. Although the role of structural disorder (zero-point motion,
thermal disorder as function of temperature and possible disorder because of Si/Ge substitutions) 
merits a more careful analyse, we may include
some of the effects from zero-point motion through the Fermi-Dirac function.
All calculations are made
with the same internal structural parameters of the B20 structure and with no 
structural disorder. Thus, only the effect of substitutional disorder appears in 
these calculations, although it is probable that different relaxation
around Si and Ge will increase the effects of disorder in real FeSi$_{(1-x)}$Ge$_x$.

The calculations do not determine the equilibrium
volumes through minimization of the total energies for each composition, but the calculated pressures (P)
are used as a guidance for finding the volumes relative to that of pure FeSi. Absolute values of direct
calculations of the pressure are less reliable than when P is calculated through the volume derivative
of total energies, but they converge more rapidly and can be
useful for studies of relative variations.
The  bands of pure FeSi at T=0 (no disorder) have a gap of 6 mRy at the Fermi energy, E$_F$, in agreement with
other calculations using different methods 
\cite{mat}-\cite{gre}. This is for a lattice constant, $a_0$=4.39  \AA, 
between the experimental one (4.52 \AA) and the theoretical one (at the minimum of the total energy).
 This reflects the usual problem of
using the LDA for 3d transition metals, in which
the lattice constants come out 2-3 percent too small
compared to experiment \cite{bmj}. 
 This uncertainty could be relatively severe for FeSi because of the narrow gap.
However, the theoretical E$_g$ agrees well with experiment and
it decreases only slowly when the
 lattice constant increases, in agreement with
the pressure dependence of the magnetic susceptibility \cite{jar,gre}.

The first calculation for FeGe is made with $a_0 =$4.52 \AA, i.e. about 3 percent larger than for FeSi. 
This choice is based on the differences in covalent radii and lattice constants of pure Ge and Si,
and as for FeSi we consider the LDA-bands for a lattice constant which is a few percent smaller than
the experimental value (4.7 \AA).
The calculated gap in FeGe, 1.5 mRy, is close to the LDA result of ref \cite{ani2}. 
For intermediate compositions 
$a_0$ is assumed to vary linearly with $x$, which turns out to be consistent with the 
calculated variations of the pressure. 
The discontinuities
of the potential at the limits of the different Wigner-Seitz spheres are small when 
the Wigner-Seitz radii are
1.44 \AA~ for Fe, and
1.37 \AA~ for Ge and 1.27 \AA~ for Si in FeGe and FeSi respectively. 
Almost the same values are
maintained in the calculations for the FeSi$_{(1-x)}$Ge$_x$-systems.
The s- and p-bands of Si and Ge are similar, both concerning the position relative to the
Fe bands and the band widths. The unoccupied 3d band in Si is almost 0.5 Ry closer to E$_F$
than the 4d band in Ge. This makes a difference for the hybridization with the Fe-d
bands near E$_F$ and for the degree of disorder of the band structure in the alloys.

Calculations are made for pure Fe$_{32}$Si$_{32}$ 
and Fe$_{32}$Ge$_{32}$ and for configurations with 4, 7, 10, 11, 12, 14, 16, 17, 19, 22, 25, 28, 30 
and 31 number of Ge sites per supercell.  
The distribution of Ge vs. Si sites is random in the different cells for different $x$,
but two calculations are made in which the Ge sites
(4 and 12, respectively) are clustered together within the cell.  Two 
calculations are made for pure FeGe, labeled II and III, 
in which $a_0$ is increased by one and two percent, respectively. As will be dicussed later,
the results for the moment agree best with experiment in calculation FeGe-III, although the
lattice constant is smaller than the experimental one.

\vspace{0.2cm}
\section{Results.}
\vspace{0.3cm}
\subsection{Bandstructure near the gap.}
\vspace{0.3cm}

The width of the gap, the unpolarized DOS at E$_F$, $N_{para}$, 
and the exchange splittings, $\xi$, from the spinpolarized calculations are shown in Table I. 
The gap, $E_g$, is about 2.5 mRy in both
calculations with 4 Ge atoms, either randomly distributed or 
clustered within one basic cell of FeGe together with 7 basic cells of FeSi.
The effect on the gap from Ge substitutions is relatively weak compared to 
other dopants replacing Fe or Si \cite{jar01}.
As might be expected, the effects of disorder on the DOS are to reduce $E_g$ 
and to make the DOS peaks near the gap wider.  
The gap is reduced when the concentration of
Ge is increased, and the alloy becomes metallic at about 8-9 Ge sites per cell.
The Fermi energy is near the bottom of a 'valley', or pseudogap, in the DOS for all paramagnetic
calculations, showing that
the gap is never completely washed out.  A real gap is restored for the two
highest Ge concentrations when only one or two Si remain in the cell.

The Fe d-bands contribute most to the large DOS near the gap, 
with some local variations from site to site due to the differences of near neighbor atoms.
Magnetic, Stoner-type ordering can start at sites with the largest $N(E_F)$. 
The magnetic instability occurs when $N(E_F) \cdot I \geq 1$, 
where $I$ is the exchange integral, but the size of the moments
is determined by other parameters. (As seen in Table I, the calculation with 12 clustered Ge atoms
has a higher paramagnetic $N(E_F)$ than the case with random site occupation, but for $\xi$
and the moment it is the other way around.)
Typically there is more than a factor of two between the lowest and highest 
local moments on different Fe when disorder plays a role,
i.e. for $x$ not too close to 1. 
The difference in local moments leads to an additional disorder
in the spin-polarized part of the potential, so that the majority and minority DOS functions are not exactly
like a rigid splitting of the paramagnetic DOS. This makes the remaining pseudogaps in the DOS of the two spins 
more 'washed out' than in the paramagnetic DOS. The mechanism leading to magnetism is accelerated as soon
as the exchange splitting allows for $E_F$ to enter into the high peaks 
of the DOS of majority and minority spin above and
below the gap, respectively. This process slows down the self-consistent convergence  
in some cases. As will be shown later, it also provides an explanation for
 the large variation of the moment in FeGe as function of volume.

The spinpolarized $N(E_F)$ and magnetic moments $m$
 are summarized in Figs. 2 and 3.
 Calculations for 4 and 7 Ge per cell give vanishing moments, while those
for 30 and 31 Ge have finite moments despite a small gap. The latter can be understood from
the thermal smearing that overcomes gaps smaller than about 1.5 mRy.
These results show that weak magnetism can be the result of substitutional disorder
of the FeSi$_{(1-x)}$Ge$_x$-system for $x$ larger than $\sim 0.3$. This range for ferromagnetism
agrees well with experiment \cite{yeo}, but the moments are smaller, and as will be discussed later,
the moments are sensitive to volume.

\vspace{0.3cm}
\subsection{Specific heat.}
\vspace{0.3cm} 

The DOS near the gap is very peaked, and statistical fluctuations of
$N(E_F)$ can be seen for the different configurations. 
The total, spin-polarized $N(E_F)$ values shown in Fig. 2 are scattered around 
800 states/Ry/cell for large Ge concentrations,  while the maximal
values (1000-1200 states/Ry/cell) are found for $x \approx 0.5$. 
The electron-phonon coupling $\lambda$ is calculated to be about 0.2 for disordered FeSi having a 
DOS of 700 states/Ry/cell \cite{jar99}. As $\lambda$ is proportional to $N(E_F)$ one can 
estimate the electronic specific heat coefficient $\gamma = \frac{1}{3} \pi^2 k_B^2 N(E_F) (1+\lambda)$ to be
$\approx$ 5-9 mJ/mole K$^2$ when $N(E_F)$ varies between 800 and 1200 states/Ry/cell.
This is lower than measured \cite{yeo}. Such discrepancies are often attributed to
 spin fluctuations that contribute to $\gamma$ through $\lambda_{sf}$ in near magnetic
systems. 
A calculation of $\lambda_{sf}$ for a single 8-atom cell of disordered FeSi (which becomes 
metallic with a moderate DOS below the limit for Stoner magnetism) indicates that it can be large, 
between 0.5 and 1. This DOS is comparable or lower than the DOS in FM phases of FeSi$_x$Ge$_{(1-x)}$,
but $\lambda_{sf}$ is often lower on the magnetic side
of a FM transition \cite{fe}. Therefore it can be expected that spinfluctuations contribute to
$\gamma$ just at the very beginning of FM and less when magnetism becomes stable at larger $x$.
The measured $\gamma$ is peaked at $\sim$ 20 mJ/mole K$^2$ near $x=0.37$ \cite{yeo}, i.e.
rather close to the critical $x$ for the metallic transition. 
From the DOS of Fig 1 it is possible to
understand this behavior. All DOS functions show the peaks around the gap or pseudogap
independently of $x$. Larger
disorder will smear the DOS, but the peak positions remain even in the spin
polarized majority and minority DOS functions. An inspection
of the DOS functions shows that a moderately large moment of 3 $\mu_B$ per 64 atom cell
(corresponding to a $\xi$ of about 9 mRy)
will have $E_F$ on the first peak above the gap within the majority DOS, and on the second peak below the gap
within the minority DOS, cf. Fig 1. Thus the total $N(E_F)$ is large
for a moment of this size, from Fig. 1 one can estimate that the combined $N(E_F)$ (majority plus minority)
is about 1200 states/cell/Ry.
If the moment is larger, then $E_F$ is found further to the right on the
majority DOS and further to the left on the minority DOS. $N(E_F)$ are lower for both spins, and
the combined $N(E_F)$ taken from
Fig. 1 is 700-800 states/cell/Ry for a large moment (7-8 $\mu_B$/cell)
and a large $\xi$ of 12-15 mRy. 
This discussion can be exemplified by rigid-band spin splittings
on the paramagnetic DOS, as in Fig. 4 for three different compositions. The total charge is conserved, whereas
a moment '$m$' is obtained from the differences in the number of occupied spins. If $m$ shows a linear increase
as function of $x$ it is expected that $\gamma(x)$ has the same shape as in Fig. 4.

The trend of increasing moments towards large Ge concentration is evident from the points in Fig. 3. 
The calculated moments are typically  2-3.5 $\mu_B$ for $x$ in the range 0.35-0.7,
while for larger Ge concentrations the moments can reach 5 $\mu_B$ or more. 
Thus, the former range of concentration
corresponds to a large $N(E_F)$ (and $\gamma$), while in the latter $N(E_F)$ is reduced. (The relative
variations in $\gamma$ will be enhanced over those in $N(E_F)$ because of the coupling factors $\lambda$.)
 This behavior is partly
confirmed in Fig. 2; $N(E_F)$ is largest for $x$ between 0.4 and 0.6, although there is scattering 
among the points due to the small set of configurations. The relative variations of $m(x)$ and $\gamma(x)$ 
fit reasonably well to
the measurements by Yeo ${\it et~al}$, but
the moments are too small when $x \rightarrow 1$.
 The measured moment at $x$=0.37, about 0.1
$\mu_B$/f.u. (equal to $\sim 3.2 \mu_B$ for the 64 atom cell), is in fair agreement with the calculations. But
 the moments for $x \rightarrow 1$, $\sim$0.4 $\mu_B$/f.u. ($\sim 12 \mu_B$ per 64 atom cell) 
\cite{yeo}, are much larger than in the calculations, and they are not yet the saturation moments.

\vspace{0.3cm}
\subsection{Properties at Ge-rich compositions.}
\vspace{0.3cm}

The calculations for $x \rightarrow 1$ indicate that a small gap
will reappear in FeGe if spin-polarization could be prevented. Calculations for undoped FeGe
are easier as they can start from results for the smaller 8-atom cells, and a few volumes
have been studied.
 The calculated pressure
for FeGe (FeGe-I at $a_0$=4.52 \AA) is about 0.1 Mbar larger than for FeSi (at $a_0$=4.39 \AA), which gives a hint
that the calculations for $x \rightarrow 1$ are made at too small lattice constants. 
In a calculation (labeled FeGe-II in Table I), in which $a_0$ is increased to 4.56 \AA~, 
the difference in pressure is reduced to about 0.05 MBar, 
and for FeGe-III when $a_0$ is 4.61 \AA~ the pressure is
almost the same as for FeSi. The (paramagnetic) gap becomes smaller as $a_0$ is increased (see Table I).
It is unusual
that a gap becomes wider with applied pressure. It is more common that the gaps are between different sub-bands,
so that the gaps become narrower when the bands become wider at larger pressure. But the gap
in FeSi and FeGe is within the Fe-d band, so the gap widens together
with the band when the lattice constant is reduced \cite{jar}, although the effect is small for FeGe
as seen in Table 1. 
A narrower gap for increased lattice constant implies that a metallic transition is approaching
and the moment in the
spin-polarized calculations increases from 0.08 $\mu_B$/f.u. in FeGe-I to 0.25 and 1.06 $\mu_B$/f.u. 
for FeGe-II and III, respectively. The increase in moment as $a_0$ is increased appears very large
when there only is a small reduction of the gap. However, in addition to a narrower gap there are
sharper increases of the DOS on both sides of the gap when the volume is increased, so that more states
come closer to $E_F$. This is probably more relevant for the evolution of the magnetic moment as function of volume
than the value of the gap itself.

The strong volume dependence of $m$ is extended towards increasing Si-compositions, but not too far.
When the lattice constant is increased by 1 percent for Fe$_{32}$Si$_{14}$Ge$_{18}$ and Fe$_{32}$Si$_{7}$Ge$_{25}$,
it is found that the moment increases only in the latter case (by more than 50 percent) compared to the moments shown
in Figure 3 at the respective composition. Thus, when the gap is just closed and the system is not yet a good
metal as in the former case, there is not a strong dependence on volume. The DOS has 'tails' near the gap,
partly due to disorder. The initially small overlap between the DOS of the conduction band and the valence band
makes the system a rather poor metal, and magnetism is still hesitant.  But in the latter case, when $x$=0.78
(as well as when $x$=1) the DOS is large or increases rapidly near a narrow gap and the effect of pressure is large.

As was discussed above,
this dramatic increase of $m$ is possible when the exchange splitting is sufficiently large to make
the system clearly metallic within both spins. 
It might seem strange that a moment develops depite the (small) gap in the DOS for Ge-rich compositions,
and it could be suspected to be a metastable state.
However, convergence of the total energies for FeGe-II show that the magnetic state is indeed
the stable one with a total energy more than 30 mRy lower than the non-magnetic one. The interesting
situation of a stable non-magnetic state and a metastable magnetic state with larger total energy is
more probable for larger gaps, and the T-dependences of the two states could be
very different so that metamagnetism, large magnetoresistance and phase transitions can be imagined. 

The moment is very sensitive to volume and
 disorder, but the combined results of the pressure calculations (compared to that of FeSi) 
and the size of the calculated
moments (compared to the measured one) indicate that the lattice constant should be close to the one used
for the calculation FeGe-III, or near
4.6 \AA, for an optimal description of the properties of FeGe. The calculated moment is then close
to the saturation moment found for fields larger than 0.3 T \cite{yeo,leb,lun}, of the order 32 $\mu_B$ per
supercell.
It also follows that  
magnetism on the Ge-rich side of FeSi$_{(1-x)}$Ge$_x$ depends more on the
increased volume than on disorder. 
The strong volume dependence of $m$ suggests that thermal expansion, apart from effects due to thermal
disorder, should lead to an unusual increase of $m$ with temperature. This hypothesis is corroborated
by the measured increase of $m(T)$ by about 10 percent between low T and just below the Curie temperature
($T_C \sim$ 280 K) 
at which $m$ drops to zero, seen in the data for $x=1$ by Yeo ${\it et~al}$.
By using a typical coefficient for thermal expansion as for Fe, one arrives at an increase of the lattice
parameter of the order 0.2 percent between low-T and room temperature. The calculated increase
of $m$ as function of $a_0$ translates into an even larger increase of $m(T)$ within this temperature
interval, but this does not take into account the type of moment disorder that finally leads to
zero effective moment at $T_C$.

\vspace{0.5cm} 
\section{Conclusion.}
\vspace{0.5cm}

In conclusion, it has been shown that LDA band calculations can give an adequate description of
the FeSi$_{(1-x)}$Ge$_x$ system if disorder and changes in volume as function of $x$ are 
accounted for. The gap is small enough, so that the combined
effects of disorder and increased volume lead to zero gap for $x \approx 0.3$. The system becomes magnetic
for larger Ge concentrations, and the behavior of the electronic specific heat is directly related to the
DOS structures of the majority and minority bands. The quantitative agreement with experiment for the moments
is best when the lattice constants for Ge-rich compositions  
are 1-2 percent larger than in the results shown in Fig. 3.

Pressure experiments can be suggested from the calculated results of pure FeGe at different volume.  The delicate
balance between magnetism in metallic FeGe and absence of magnetism in semi-conducting FeGe could be followed
continuously as function of increasing pressure. 
The effect of uniform pressures would be the same as varying
$x$ from FeGe towards FeSi, but with the advantage that effects of substitutional disorder can be excluded.
Theoretical estimates of the bulk modulus (B) within LDA 
are often larger than what is found experimentally, as is the case for FeSi \cite{jar}. 
From the results above, leading to B $\approx$
1.6 Mbar, it is expected that a pressure of 0.1 Mbar will be sufficient for a suppression of the moment. 
This represents an upper limit, since the measured low value of the Debye temperature 
for FeGe \cite{yeo} is an indication of a low B. 
Information about disorder, also structural and thermal ones, is important for calculations of the
properties of the isoelectronic alloys FeSi$_{(1-x)}$Ge$_x$
because of the large DOS peaks close to a very small gap. 

Finally, the fact that properties depend on a tiny gap between high DOS peaks motivates a comment about 
results of band calculations. When the DOS near the Fermi energy is rather flat, as in most metallic materials,
there are no big consequences of details (choice of basis, linearization energy, size of atomic spheres,
general potential, type of density functional and so on) in the band theory method. But here for FeSi and FeGe, when the DOS
varies from zero to very large amplitude within 1-2 mRy, it might be that differences in the method of calculation
lead to quite different properties. It is important to note that the gap of about 6 mRy in FeSi agree with
experiments and other calculations. Also the result that FeSi$_{(1-x)}$Ge$_x$ becomes metallic for $x$ larger than
about 0.3 agrees with experiment. But there is little experimental information about the existence of a small gap (1-2 mRy)
in the spin polarized bands of pure FeGe. Therefore, experimental results of FeGe as function of pressure
would be valuable.

{\noindent {\bf Acknowledgements:}}
I am grateful to F.P. Mena, D. van der Marel and H. Wilhelm for helpful discussions.
 
%\newpage

%  
\newpage
\noindent
TABLE I. Calculated band gap and  
density-of-states at $E_F$, $N_{para}(E_F)$, in units of states/cell/Ry from paramagnetic calculations, and
exchange splitting, $\xi$, from the spin-polarized calculations. The labels "-c"
mean clustered configuration (see text), and I,II and III for pure FeGe are for lattice constants
4.52, 4.56 and 4.61 \AA.
%{\bf Table caption.}
\begin{table}[htbp]
\begin{tabular}{ccccc}
  Cell & E$_g$ (mRy) & $N_{para}$ & $\xi$ (mRy)  \\
\hline
 Fe$_{32}$Si$_{32}$ & 6 & 0 & -\\
 Fe$_{32}$Si$_{28}$Ge$_4$ & 2.5 & 0 & - \\
 Fe$_{32}$Si$_{28}$Ge$_4$-c & 2.5 & 0 & - \\
 Fe$_{32}$Si$_{25}$Ge$_7$ & 2 & 0 & - \\
 Fe$_{32}$Si$_{22}$Ge$_{10}$ & - & 16 & 1 \\
 Fe$_{32}$Si$_{21}$Ge$_{11}$ & - & 3 & 2\\
 Fe$_{32}$Si$_{20}$Ge$_{12}$ & - & 110 & 9 \\
 Fe$_{32}$Si$_{20}$Ge$_{12}$-c & - & 190 & 6\\
 Fe$_{32}$Si$_{18}$Ge$_{14}$ & - & 90 & 5\\
 Fe$_{32}$Si$_{16}$Ge$_{16}$ & - &  80 & 7\\
 Fe$_{32}$Si$_{15}$Ge$_{17}$ & - & 100 & 5\\
 Fe$_{32}$Si$_{13}$Ge$_{19}$ & - & 110 & 9\\ 
 Fe$_{32}$Si$_{10}$Ge$_{22}$ & - & 75 & 9\\
 Fe$_{32}$Si$_{7}$Ge$_{25}$ & - & 27 & 12\\
 Fe$_{32}$Si$_{4}$Ge$_{28}$ & - & 13 & 13\\
 Fe$_{32}$Si$_{2}$Ge$_{30}$ & 0.5 & 0 & 12\\
 Fe$_{32}$Si$_{1}$Ge$_{31}$ & 0.8 & 0 & 3\\
 Fe$_{32}$Ge$_{32}$-I & 1.3 & 0 & 5\\
 Fe$_{32}$Ge$_{32}$-II & 1.1 & 0 & 18\\
 Fe$_{32}$Ge$_{32}$-III & 0.8 & 0 & 78\\    
\hline
\end{tabular}
%\label{tab:tm}
\end{table}

%{\bf Figure Captions}

\begin{figure}[tb!]
%\begin{center}

% figures directly (hr bw)\begin{figure}[tb!]
\leavevmode\begin{center}\epsfxsize8.6cm\epsfbox{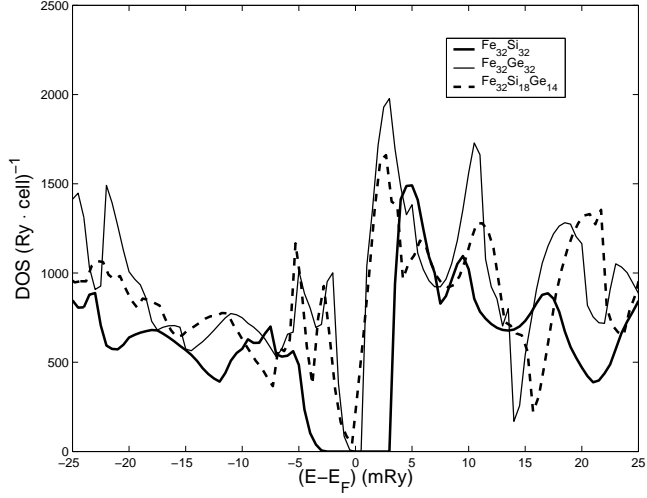}\end{center}
\caption{Paramagnetic density-of-states of FeSi, FeGe and FeSi$_{0.56}$Ge$_{0.44}$
as calculated for the supercells containing totally 64 atoms.
The energy is relative to $E_F$.
}
\end{figure}

\begin{figure}[tb!]
\leavevmode\begin{center}\epsfxsize8.6cm\epsfbox{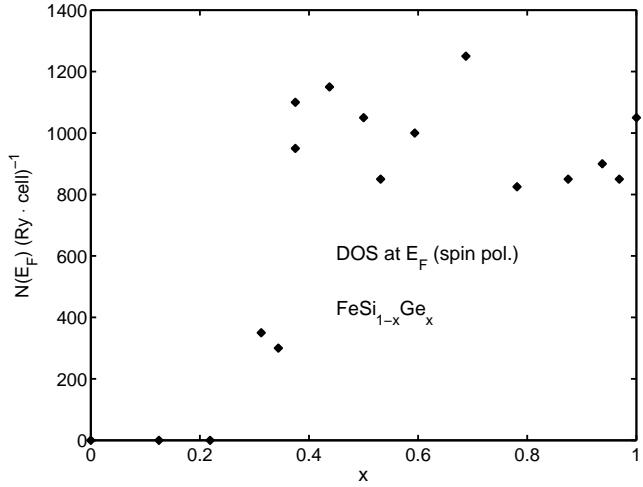}\end{center}
\caption{Spinpolarized density-of-states at the Fermi energy, $N(E_F)$, as function of the Ge
concentration $x$ in unitcells containing 64 atoms.
}
\end{figure}

\begin{figure}[tb!]
\leavevmode\begin{center}\epsfxsize8.6cm\epsfbox{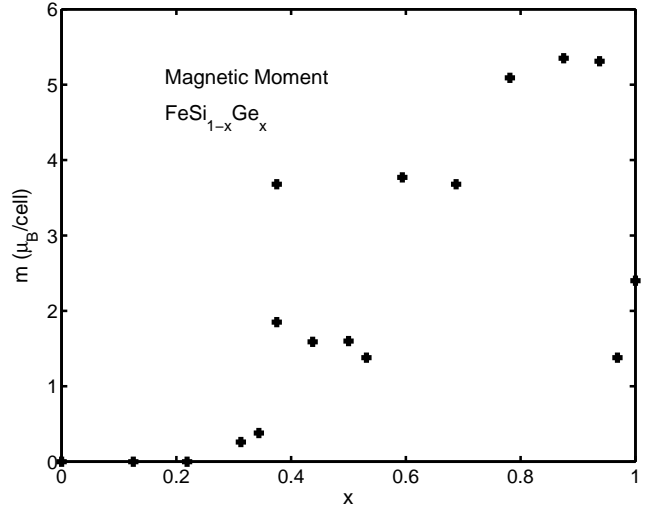}\end{center}
\caption{Magnetic moments as function of the Ge
concentration $x$ in cells containing 64 atoms. Note that the
two calculations for pure FeGe ($x$=1) for 1 and 2 percent larger 
lattice constants have much larger moments, 8.2 and 33.8 $\mu_B$/cell 
respectively. Calculations for 1 percent larger lattice constants for $x$=0.44
and $x$=0.78 give increased moment only in the latter case, near 8 $\mu_B$/cell.
}
\end{figure}

\begin{figure}[tb!]
\leavevmode\begin{center}\epsfxsize8.6cm\epsfbox{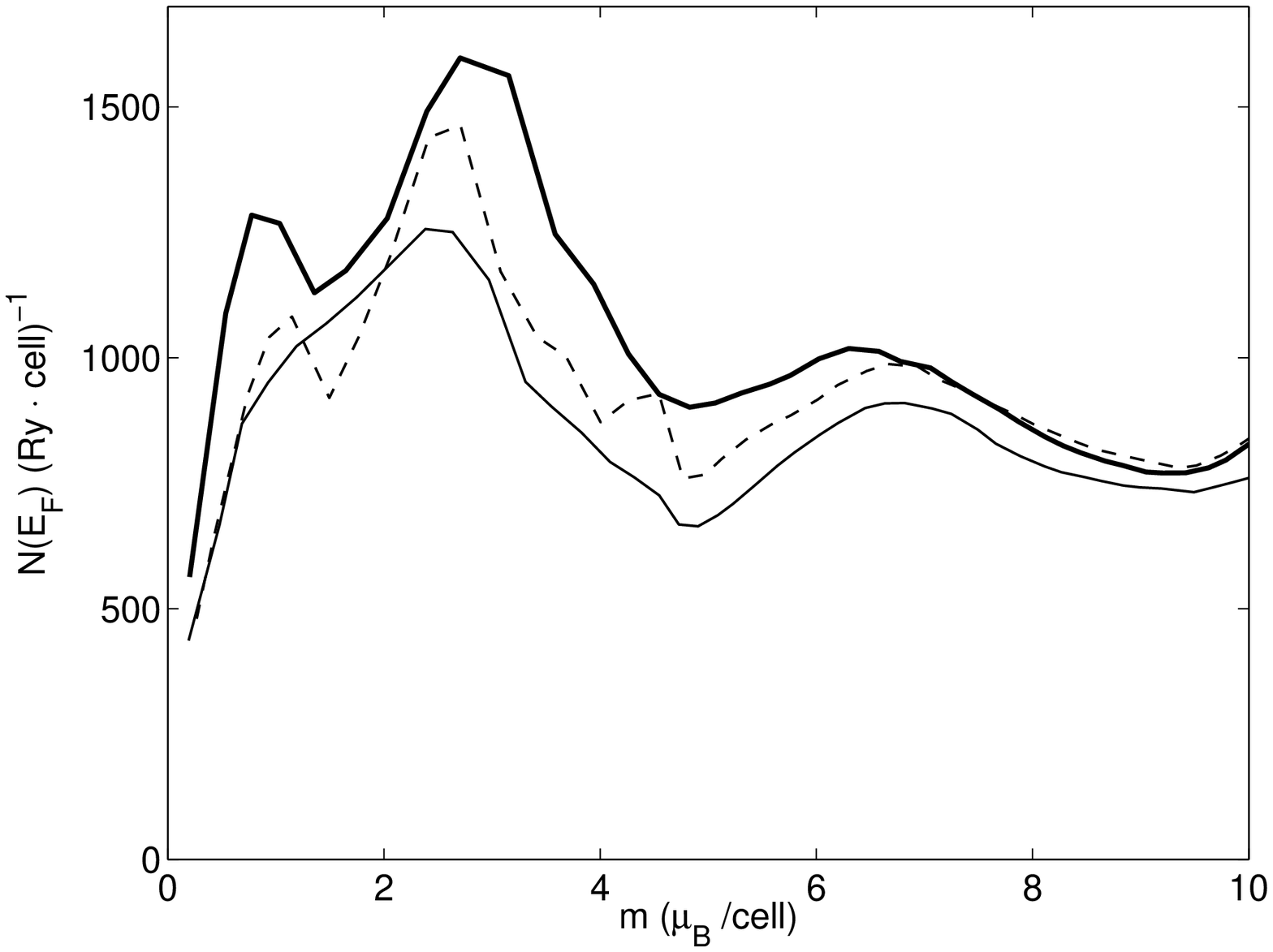}\end{center}
\caption{Total $N(E_F)$ as function of magnetic moments calculated from 
a rigid spin splitting of the paramagetic DOS functions in Fig. 1.
The heavy line is obtained when using the DOS for $x$=0, the broken line for $x$=0.44 and the thin line
for $x$=1. The variations of the observed specific heat of Yeo et al (shown in the inset of fig. 4
of ref. \cite{yeo}) show one peak near $x$=0.37, corresponding to a small  $m$.
}
\end{figure}


\begin{thebibliography}{99}
%
\bibitem{jac}
V. Jaccarino, G.K. Wertheim, J.H. Wernick, L.R.Walker and S Arajs
Phys. Rev. {\bf 160}, 476 (1967).

\bibitem{man}
D. Mandrus, J.L. Sarrao, A. Migliori, J.D. Thompson and Z. Fisk, 
Phys. Rev. B{\bf 51}, 4763 (1995).

\bibitem{sal}
B.C. Sales, E.C. Jones, B.C. Chakoumakos, J.A. Fernandez-Baca, H.E. Harmon,
J.W. Sharp and E.H. Volckmann, 
Phys. Rev. B{\bf 50}, 8207 (1994).

\bibitem{par} C-H Park, Z-H Shen, A.G. Loeser, D.S. Dessau, D.G. Mandrus,
A. Migliori, J. Sarrao and Z. Fisk, Phys. Rev. B{\bf 52}, R16981, (1995).

\bibitem{sch} Z. Schlesinger, Z. Fisk, H-T Zhang, M.B. Maple, J.F. DiTusa
and G. Aeppli, Phys. Rev. Lett. {\bf 71}, 1748 (1993).

\bibitem{deg} L. Degiorgi, M.B. Hunt, H.R. Ott, M. Dressel,
B.J. Fenstra, G. Gr\"uner, Z. Fisk and P. Canfield, Europhys. Lett. {\bf 28}, 341 (1994).

\bibitem{mor} Y. Takahashi and T. Moriya, J. Phys. Soc. Jap. {\bf 46}, 
1451 (1979);  S.N. Evangelou and D.M. Edvards, J. Phys. C{\bf 16}, 2121 (1983)

\bibitem{app} G. Aeppli and Z. Fisk, Comm. Cond. Mat. Phys. {\bf 16}, 155, (1992).

\bibitem{mat}  
L.F. Mattheiss and D.R. Hamann 
Phys. Rev. B{\bf 47}, 13114(1993). 

\bibitem{ani}
V.I. Anisimov, S.Yu Ezhov, I.S. Elfimov, I.V. Solovyev and T.M. Rice,
Phys. Rev. Lett. {\bf 76}, 1735 (1996).

\bibitem{don} C. Fu, M.P. Krijn, and S. Doniach
Phys. Rev.  B{\bf 49}, 2219 (1994) 

\bibitem{jar}
T. Jarlborg, Phys. Rev. B{\bf 51}, 11106 (1995).

\bibitem{jar99}
T. Jarlborg, Phys. Rev. B{\bf 59}, 15002 (1999);
 Phys. Lett. A{\bf 236}, 143, (1997).

\bibitem{gre}  
G.E. Grechnev, T. Jarlborg, A.S. Panfilov, M. Peter and I.V. Svechkarev, 
Solid State Commun. {\bf 91}, 835 (1994).

\bibitem{voc} L. Vo\v{c}adlo, G.D. Price and I.G. Wood, Acta Cryst. B{\bf 55}, 484, (1999).

\bibitem{lda} W. Kohn and L.J. Sham, Phys Rev {\bf 140}, A1133 (1965).

\bibitem{mena} F.P. Mena, D. van der Marel, A. Damascelli, M. F\"{a}th,
A.A. Menovsky amd J.A. Mydosh, Phys. Rev. B{\bf 67}, 241101(R), (2003).

\bibitem{jar01} T. Jarlborg, Physica B{\bf 293}, 224, (2001).

\bibitem{yeo}
S. Yeo, S. Nakatsuji, A.D. Bianchi, P. Schlottmann, Z. Fisk, L. Balicas, P.A. Stampe
and R.J. Kennedy,
Phys. Rev. Lett. {\bf 91}, 046401 (2003).

\bibitem{ani2}
V.I. Anisimov, R. Hlubina, M.A. Korotin, V.V. Mazurenko, T.M. Rice, A.O. Shorikov and M. Sigrist,
Phys. Rev. Lett. {\bf 89}, 257203 (2002).

\bibitem{leb} B. Lebech, J. Bernard and T. Frelhoft, J. Phys. Condens. Matter {\bf 1}, 6105 (1989).

\bibitem{lun} L. Lundgren, K.\AA. Blom and O. Beckman, Phys. Lett. {\bf 28A}, 175 (1968).

\bibitem{bmj} B. Barbiellini, E.G. Moroni and T. Jarlborg, J. Phys.: 
Condens. Matter {\bf 2}, 7597 (1990).
 
\bibitem{js} T. Jarlborg and G. Santi, Physica C{\bf 329}, 243, (2000).

\bibitem{jar96} T. Jarlborg, Phys. Rev. Lett. {\bf 77}, 3693 (1996).

\bibitem{fe} T. Jarlborg, Physica C{\bf 385}, 513, (2003).

%\bibitem{lmt} O.K. Andersen, Phys. Rev. B{\bf 12}, 3060 (1975);
%T. Jarlborg and G. Arbman, J. Phys. F{\bf 7}, 1635 (1977)
%\bibitem{neu} G. Shirane, J.E. Fischer, Y. Endoh and K. Tajima
%Phys. Rev. Lett. {\bf 59}, 351 (1987). 
%\bibitem{mena} F.P. Mena, D. van der Marel, C. Presura, A. Damascelli,
%G. Aeppli, J. DiTusa, A.A. Menovsky amd J.A. Mydosh, (unpublished 2003). 
%\bibitem{jf} T. Jarlborg and A.J. Freeman, Phys. Rev. B{\bf 23}, 3577 (1981).
\end{thebibliography}
\end{document}